\documentclass{article}
\usepackage{spconf,amsmath,graphicx,hyperref}

\usepackage{cite}
\usepackage{overpic}
\usepackage{amsmath,amssymb,amsfonts}
\usepackage{graphicx}
\usepackage{textcomp}
\usepackage{xcolor}
\usepackage{float}
\usepackage{amsthm}
\usepackage{graphicx}
\usepackage{epstopdf}
\usepackage{amsmath,bm}
\usepackage{amsfonts}
\usepackage{amssymb}
\usepackage{color}
\usepackage{subfigure}
\usepackage{multirow}
\usepackage{multicol}
\usepackage{url}
\usepackage{soul,xcolor}
\usepackage{algorithm}
\usepackage{algpseudocode}%

\theoremstyle{plain}

\newcommand{\vect}[1]{\mathbf{#1}}

\def\diag{\mathrm{diag}}

\def\Htran{\mbox{\tiny $\mathrm{H}$}}
\def\Ttran{\mbox{\tiny $\mathrm{T}$}}
\def\CN{\mathcal{N}_{\mathbb{C}}} 
\def\imagunit{\mathsf{j}}

\def\sinc{\mathrm{sinc}}

\def\imaginary{\mathsf{j}} 
\def\imagunit{\imaginary} 

\def\sinc{\mathrm{sinc}}

\def\pb{\textrm{p}}

\title{Capacity Analysis of OFDM Systems with a Swarm of Network-Controlled Repeaters}

\name{Doğa Evgür$^*$, Ozan Alp Topal$^\dagger$, and Özlem Tuğfe Demir$^\ddagger$
}
\address{$^*$Electrical-Electronics Engineering, TOBB University of Economics and Technology, Ankara, Turkiye \\
$^\dagger$Department of Communication Systems, KTH Royal Institute of Technology, Stockholm, Sweden \\
$^\ddagger$Department of Electrical and Electronics Engineering, Bilkent University, Ankara, Turkiye \\
Email: devgur@etu.edu.tr, oatopal@kth.se, ozlemtugfedemir@bilkent.edu.tr
}

\begin{document}
\ninept

\maketitle

\begin{abstract}
This paper investigates the uplink capacity of single-input single-output (SISO) systems assisted by a swarm of network-controlled repeaters (NCRs). We develop a rigorous wideband formulation based on OFDM signaling. Starting from the continuous-time passband model, we derive the capacity expression for the repeater-assisted OFDM channel, accounting for amplified noise contributions from multiple repeaters. Numerical results demonstrate that NCRs can substantially enhance system capacity even with simple activation strategies, and that activating only the closest repeater yields nearly the same performance as activating all repeaters, thereby offering significant energy-saving opportunities. These findings highlight the potential of NCR swarms as a cost-effective and scalable solution for coverage extension and capacity enhancement in wideband wireless networks.

\end{abstract}


\vspace{-2mm}

\section{Introduction}
\label{sec:intro}
\vspace{-2mm}

Network-controlled repeaters (NCRs) are a specific type of repeaters that can be controlled by a central node. They are capable of amplifying and forwarding the received signals within nanoseconds of delays, consequently acting as active scatterers by effectively increasing the number of received paths~\cite{willhammar2025achieving}. NCRs have been extensively used when coverage extension is required, and are already part of Release 18 \cite{wen2024shaping}. NCR is a low-cost alternative to the deployment of additional base stations (BSs) or reconfigurable intelligent surfaces (RISs). Unlike BS or RIS, NCR does not require complex channel state information (CSI) acquisition algorithms or control in every coherence block, effectively reducing the fronthaul and computational requirements and hardware cost. 

Due to these advantages, NCRs have recently been envisioned not just for coverage extension but also to increase the macro diversity similar to distributed multi-input-multi-output (D-MIMO) systems by acting as active scatterers with amplification. This can be satisfied by distributing a swarm of NCRs in the cell, and controlling them by the BS \cite{willhammar2025achieving}, \cite{bai2025repeater}. The main difference between D-MIMO systems is the lack of processing capability, limited fronthaul, and control capability in the NCRs. Due to the lack of processing capability, NCRs amplify the noise within their received signal, where in a system with a swarm of NCRs, some NCRs might act as noise sources due to the high distance between the user equipment (UE) and NCRs. Therefore, amplification control for NCRs is necessary to unlock the true potential of such a system. The authors of \cite{topal2025fair} demonstrate that frequent control at the coherence block level is unnecessary; instead, controlling them at the second level with quantized amplification control signaling provides similar capacity gains. In \cite{larsson2024stability}, the amplification limits for stability in NCRs are provided to ensure that repeaters do not continuously amplify each other's signals. 

However, all the works listed above rely on a simplified narrowband signal assumption, overlooking the frequency-selective behavior of wideband characteristics, and resulting in overly positive capacity gains. To fill this research gap, we model the capacity of an uplink channel with a swarm of NCRs, considering a single-input-single-output (SISO) OFDM system. The main contributions of this paper are summarized as follows.  
\begin{itemize}
    \item We develop a rigorous wideband system model for repeater-assisted SISO-OFDM, starting from the continuous-time passband representation. Unlike prior narrowband approximations, our formulation explicitly incorporates frequency-selective propagation and noise amplification effects.  
    \item By discretizing the channel into multipath components, we derive closed-form expressions for the effective channel coefficients, where the contributions of the direct path and all repeaters are separated into different terms, which highlights the role of the amplification factors of the repeaters.  
    \item We analyze the noise statistics rigorously by deriving the covariance structure of the amplified noise across subcarriers.  Based on this model, we establish the capacity expression of the repeater-assisted OFDM channel in terms of the singular values of the whitened channel matrix, combined with water-filling power allocation. This provides the first mathematically exact wideband capacity formulation for NCR swarms.  
\end{itemize}

\vspace{-2mm}
\section{System Model}
\vspace{-2mm}

We consider the uplink of a repeater-assisted SISO OFDM system consisting of a single-antenna BS, a single-antenna UE, and $L$ single-antenna repeaters. The repeaters can be arbitrarily located in the considered coverage area. As in \cite{willhammar2025achieving}, for simplicity, we will disregard the interactions between the repeaters. We will mainly use the notation adopted in \cite{bjornson2024introduction,demir2024wideband}. We denote the continuous-time passband signal emitted by the UE as $x_{\pb}(t)$ and let  $g_{\textrm{d},\pb}(t)$ denote the impulse response of the linear time-invariant direct channel between UE and BS, excluding the paths to/from the repeaters. Additionally, $g_{\textrm{ue},l,\pb}(t)$ and $g_{\textrm{bs},l,\pb}(t)$ denote the impulse responses of the channels from the UE to repeater $l$ and from that repeater to BS, respectively. Therefore, the received signal at repeater $l$ can be expressed as 
\begin{align} 
\tilde{r}_{\pb}(t) =&\left(g_{\textrm{ue},l,\pb} *x_{\pb}\right)(t) +w_{l,\pb}(t)
\end{align}
where $w_{l,\pb}(t)$ is the white circularly symmetric complex Gaussian random process with constant power spectral density $N_0$\,W/Hz. The repeater $l$ will amplify and transmit the received signal with the amplification factor $\alpha_l>0$ and a certain small delay $\tau_l$.\footnote{The repeater delay is on the order of a few nanoseconds; therefore, this delay does not significantly increase the cyclic prefix length.} The impulse response of the repeater $l$ is then denoted by $\vartheta_{l,\pb}(t)=\alpha_l\delta(t-\tau_l)$.

The received signal at the BS through the direct channel and the repeaters is given by
\begin{align} 
\upsilon_{\pb}(t) 
&=(g_{\textrm{d},\pb}
* x_{\pb})(t)\nonumber\\
&\quad + \sum_{l=1}^{L}\left( g_{\textrm{bs},l,\pb}*\vartheta_{l,\pb}*g_{\textrm{ue},l,\pb} *x_{\pb}\right)(t) \nonumber \\
&\quad + \sum_{l=1}^{L}\left( g_{\textrm{bs},l,\pb}*\vartheta_{l,\pb}*w_{l,\pb}\right)(t) +w_{\textrm{bs},\pb}(t) 
\end{align}
where $w_{\textrm{bs},\pb}(t)$ is the white circularly symmetric complex Gaussian random process with constant power spectral density $N_0$\,W/Hz. Next, we define the passband end-to-end impulse response of the channel as
\begin{align} 
g_{\pb}(t) &=
g_{\textrm{d},\pb}(t)
+ \sum_{l=1}^L( g_{\textrm{bs},l,\pb}*\vartheta_{l,\pb}*g_{\textrm{ue},l,\pb})(t). \label{eq:end-to-end}
\end{align}
The received signal at the BS can then be written as
\begin{align} 
\upsilon_{\pb}(t)  &=(g_{\pb}* x_{\pb})(t) \nonumber \\&\quad+ \sum_{l=1}^{L}\left( g_{\textrm{bs},l,\pb}*\vartheta_{l,\pb}*w_{l,\pb}\right)(t) +w_{\textrm{bs},\pb}(t).
\end{align}

By down-shifting all the impulse responses  shown in \eqref{eq:end-to-end} and noise process for the carrier frequency of $f_{\mathrm{c}}$ as
\begin{align}
&g_{\textrm{d}}(t) = g_{\textrm{d},\pb}(t) e^{-\imagunit 2\pi f_{\mathrm{c}} t}, \\
& g_{\textrm{ue},l}(t) = g_{\textrm{ue},l,\pb} (t)e^{-\imagunit 2\pi f_{\mathrm{c}} t}, \\
& g_{\textrm{bs},l} (t) = g_{\textrm{bs},l,\pb} (t) e^{-\imagunit 2\pi f_{\mathrm{c}} t}, \\
& \vartheta_{l}(t) = \vartheta_{l,\pb} (t) e^{-\imagunit 2\pi f_{\mathrm{c}} t}, \\
&w_{l}(t) = w_{l,\pb}(t)e^{-\imagunit 2\pi f_{\mathrm{c}} t}
\end{align}
 we can obtain the complex baseband representation of the end-to-end impulse response of the channel
\begin{align} 
g(t) &=
g_{\textrm{d}}(t)
+ \sum_{l=1}^L(g_{\textrm{bs},l}*\vartheta_{l}*g_{\textrm{ue},l})(t).
\end{align}
The transmitted passband signal $x_\pb(t)$ can be expressed in complex baseband form as $x(t)$ which can be generated using pulse amplitude modulation (PAM) given by  $x(t) = \sum_{k=-\infty}^{\infty}x[k]p\left(t-\frac{k}{B}\right)$, where $\{x[k]\}$ denotes the discrete sequence of data symbols transmitted at the symbol rate $B$. $p(t)$ is the ideal sinc pulse  $p(t) = \sqrt{B} \, \text{sinc}(Bt)$  and $\sqrt{B}$ normalizes the sinc pulse to have unit energy. 

When the received complex baseband signal is lowpass filtered with the pulse $p(t)$, and sampled at the symbol rate $B$, the resulting signal at the BS becomes
\begin{align}
y[r] = \sum_{\ell=0}^{T} h[\ell]  x[r-\ell]  + n[r],  \label{eq:received-time}
\end{align}
where the channel is discretized into $T + 1$ coefficients, where $T$ is chosen such that $h[\ell]\approx 0$ for $\ell>T$ and $n[r]$ is the effective noise due to the superposition of the amplified noise by the repeaters and receiver antenna noise. 

We first elaborate on the channel coefficients, which  are given  by
\begin{align} \nonumber
h[\ell] &= (p * g * p) \left(  t \right) \Big|_{t=\ell/B} =  (p * g_{\textrm{d}} * p) (t) \Big|_{t=\ell/B}  \nonumber \\
 &\quad  + \sum_{l=1}^{L} (p * g_{\textrm{bs},l} * \vartheta_{l} * g_{\textrm{ue},l} * p) (t) \Big|_{t=\ell/B}.   \label{eq:discrete}
\end{align}

The complex baseband impulse response of the repeater $l$ is given as  $ \vartheta_{l} (t) = \alpha_l\delta(t  - \tau_{l}) e^{-\imagunit 2\pi f_{\mathrm{c}} t}$. The impulse response of the direct channel between the UE and the BS contains $L_{\textrm{d}}$ propagation paths and is expressed as
\begin{align} 
 \quad g_{\textrm{d}}(t) = \sum_{i=1}^{L_{\textrm{d}}} \alpha_{\textrm{d},i}  e^{-\imagunit 2\pi f_{\mathrm{c}} t} \delta( t + \eta - \tau_{\textrm{d},i})
\end{align}
where the subscript $i$ indexes $i$th propagation path. The parameter $\alpha_{\textrm{d},i} \in [0,1]$ denotes the attenuation coefficient, $\tau_{\textrm{d},i} \geq 0$ represents the corresponding path delay and the common receiver clock offset at the BS receiver is given by $\eta$. 

The impulse responses of the channels from the UE to repeater $l$ and from that repeater to the BS are similarly written as  
\begin{align}
g_{\textrm{ue},l}(t) &= \sum_{i=1}^{L_{\textrm{ue},l}} \alpha_{\textrm{ue},l,i} e^{-\imagunit 2\pi f_{\textrm{c}} t} \delta(t - \tau_{\textrm{ue},l,i})\\
g_{\textrm{bs},l}(t) &= \sum_{j=1}^{L_{\textrm{bs},l}} \alpha_{\textrm{bs},l,j} e^{-\imagunit 2\pi f_{\textrm{c}} t} \delta(t + \eta - \tau_{\textrm{bs},l,j}). 
\end{align}
The constants $L_{\textrm{ue},l}$ and $L_{\textrm{bs},l}$ represent the number of propagation paths in the UE-repeater and repeater-BS links, respectively.  $ \alpha_{\textrm{ue},l,i},  \alpha_{\textrm{bs},l,j}\in [0,1]$ are attenuation coefficients, and the corresponding path delays are given by $ \tau_{\textrm{ue},l,i}, \tau_{\textrm{bs},l,j}\geq 0$.  

We then rearrange \eqref{eq:discrete} by means of the properties $(p * p)(t)= \sinc ( Bt )$ and $\sinc ( Bt )*e^{-\imagunit 2\pi f_{\textrm{c}} t} \delta(t - \tau)=\sinc ( B(t - \tau))e^{-\imagunit 2\pi f_{\textrm{c}} \tau} $.  This yields the expression
\begin{align} \nonumber
&h[\ell] = 
\sum_{i=1}^{L_{\textrm{d}}}  \alpha_{\textrm{d},i}  e^{-\imagunit 2\pi f_{\mathrm{c}} (\tau_{\textrm{d},i}-\eta)}
\sinc \big( \ell + B( \eta -\tau_{\textrm{d},i}) \big) +\\
& \sum_{l=1}^{L} \sum_{i=1}^{L_{\textrm{ue},l}} \sum_{j=1}^{L_{\textrm{bs},l}} \alpha_{\textrm{ue},l,i} \alpha_{\textrm{bs},l,j}\alpha_{l}  e^{-\imagunit 2\pi f_{\mathrm{c}} (\tau_{\textrm{ue},l,i}+\tau_{\textrm{bs},l,j}+\tau_{l} - \eta)}
\nonumber\\
&\times\sinc \big( \ell + B( \eta -\tau_{\textrm{ue},l,i}-\tau_{\textrm{bs},l,j}- \tau_{l}) \big).
\end{align}

By rearranging the channel coefficients to isolate the amplification factors of all repeaters, we obtain 
\begin{align}
c_{\textrm{d}}[\ell]  &=   \sum_{i=1}^{L_{\textrm{d}}}  \alpha_{\textrm{d},i}  e^{-\imagunit 2\pi f_{\mathrm{c}} (\tau_{\textrm{d},i}-\eta)}
\sinc \big( \ell + B( \eta -\tau_{\textrm{d},i}) \big), \label{eq:direct-last} \\
c_{\textrm{r},l}[\ell] &= \sum_{i=1}^{L_{\textrm{ue},l}} \sum_{j=1}^{L_{\textrm{bs},l}} \alpha_{\textrm{ue},l,i} \alpha_{\textrm{bs},l,j}  e^{-\imagunit 2\pi f_{\mathrm{c}} (\tau_{\textrm{ue},l,i}+\tau_{\textrm{bs},l,j}+\tau_{l} - \eta)}
\nonumber\\
&\quad \times\sinc \big( \ell + B( \eta -\tau_{\textrm{ue},l,i}-\tau_{\textrm{bs},l,j}-\tau_l) \big). \label{eq:repeater-channel-time}
\end{align}
We vectorize \eqref{eq:repeater-channel-time} with respect to the repeater index $l$ and obtain 
\begin{align}
\vect{c}_{\textrm{r}}[\ell]=&
\begin{bmatrix}
c_{\textrm{r},1}[\ell] \\
c_{\textrm{r},2}[\ell]\\
\vdots \\
c_{\textrm{r},L}[\ell]
\end{bmatrix} \in \mathbb{C}^{L}. \label{eq:vector-cr}
\end{align}
 Using \eqref{eq:vector-cr}, we can express the channel coefficients as
\begin{align}
h[\ell]&=
c_{\textrm{d}}[\ell]+\vect{c}_{\textrm{r}}^{\Ttran}[\ell]\bm{\alpha}
\end{align}
where, $\bm{\alpha}=
\begin{bmatrix}
\alpha_{1} & \alpha_{2} & \cdots & \alpha_{L} 
\end{bmatrix}^{\Ttran} \in \mathbb{R}^{L}$.

By applying OFDM modulation with $S$ data symbols per block and a cyclic prefix of length $T$, which is smaller than $S$, we obtain $S$ subcarriers of the form:
\begin{align} 
\bar{y}[\nu] =\bar{h}[\nu] \bar{\chi}[\nu] + \bar{n}[\nu], \quad \textrm{for} \,\, \nu = 0, \ldots, S-1, \label{eq:frequency}
\end{align}
where $ \bar{\chi}[\nu]$ indicates the transmitted data signal at the subcarrier $\nu$ as described in \cite[Sec. 7.1.1]{bjornson2024introduction} and $\bar{n}[\nu]$ is the independent noise whose statistics will be derived later. Defining $\bar{c}_{\textrm{d}}[\nu] = \sum_{\ell = 0}^{T} c_{\textrm{d}}[\ell]  e^{-\imaginary 2 \pi \ell \nu /S}$ and  $\bar{\vect{c}}_{\textrm{r}}[\nu] = \sum_{\ell = 0}^{T} \vect{c}_{\textrm{r}}[\ell] e^{-\imaginary 2 \pi \ell \nu /S}$, for $\nu=0,\ldots,S-1$, we obtain the frequency domain channel coefficients as:
\begin{align}
\bar{h}[\nu]&=
\bar{c}_{\textrm{d}}[\nu]+\bar{\vect{c}}_{\textrm{r}}^{\Ttran}[\nu]\bm{\alpha}.
\end{align}

We now analyze the statistics of the effective noise from \eqref{eq:received-time}. The time-domain effective noise signal, $n[r]$, is given as
\begin{align} \nonumber
n[r] &= \sum_{l=1}^{L} \underbrace{(p * g_{\textrm{bs},l} * \vartheta_{l} * w_l) (t) \Big|_{t=r/B}}_{\triangleq n_{l}^\prime[r]}\nonumber\\
&\quad+\underbrace{(p*w_{\textrm{bs}})(t)\Big|_{t=r/B}}_{\triangleq n^{\prime\prime}[r]}.   
\end{align}
Since $w_l(t)$ and $w_{\textrm{bs}}(t)$  are independent Gaussian processes, $n_{l}^\prime[r]$ and $n^{\prime\prime}[r]$ are independent of each other. The latter signal, $n^{\prime\prime}[r]$, has the distribution $\CN(0,N_0)$ \cite[Eq. (2.123)]{bjornson2024introduction}. Under the deterministic channel assumption,  $n_{l}^\prime[r]$ is also complex Gaussian whose variance and covariance will be derived as follows. To this end, we define the channel 
\begin{align}
g'_l(t) & = (p*g_{\textrm{bs},l}* \vartheta_{l})(t) \nonumber\\
&= \sum_{j=1}^{L_{\textrm{bs},l}}  \alpha_{\textrm{bs},l,j}  e^{-\imagunit 2\pi f_{\mathrm{c}} (\tau_{\textrm{bs},l,j}+\tau_{l} - \eta)}
\nonumber\\
&\quad \times\sqrt{B}\sinc \big( B(t + \eta-\tau_{\textrm{bs},l,j}- \tau_{l}) \big)
\end{align}
which leads to
\begin{align}
n_{l}^\prime[r] = \alpha_l (g_l'*w_l)(t)\Big|_{t=r/B}.
\end{align}
Then, the covariance between $n_{l}^\prime[r_1]$ and $n_{l}^\prime[r_2]$ can be computed using whiteness of $w_l(t)$ as
\begin{align}
\mathbb{E}\left\{n_{l}^\prime[r_1]n_{l}^{\prime*}[r_2]\right\}=\alpha_l^2\underbrace{N_0\int_{-\infty}^{\infty} g_l'\left(\frac{r_1}{B}-\tau\right)g_l^{\prime*}\left(\frac{r_2}{B}-\tau\right)d\tau}_{\triangleq d_{l,r_1,r_2}}. \label{eq:covariance}
\end{align}
Using the identity 
\begin{align}
&\int_{-\infty}^{\infty} \sinc\left(B\left(t_1-\tau\right)\right)\sinc\left(B\left(t_2-\tau\right)\right)d\tau \nonumber \\
&= \frac{1}{B} \sinc\left(B\left(t_1-t_2\right)\right),
\end{align}
we can express $d_{l,r_1,r_2}$ in \eqref{eq:covariance} as
\begin{align}
d_{l,r_1,r_2} &= N_0 \sum_{i=1}^{L_{\textrm{bs},l}}\sum_{j=1}^{L_{\textrm{bs},l}}  \alpha_{\textrm{bs},l,i}\alpha_{\textrm{bs},l,j}  e^{-\imagunit 2\pi f_{\mathrm{c}} (\tau_{\textrm{bs},l,i}-\tau_{\textrm{bs},l,j})}
\nonumber\\
&\quad \times\sinc \big(r_1-r_2- B(\tau_{\textrm{bs},l,i} -\tau_{\textrm{bs},l,j}) \big).
\end{align}
Collecting the time-domain noise entries in a vector as
\begin{align}
\vect{n}= \begin{bmatrix} n[0] \\ n[1] \\ \vdots \\ n[S-1]\end{bmatrix}
\end{align}
and using the independence of $w_l(t)$ for different $l$, we obtain $\vect{n}\sim \CN(\vect{0},\vect{D})$, where
the covariance matrix is given as 
\begin{align}
    \vect{D}= \sum_{l=1}^L \alpha_l^2\vect{D}_l + N_0\vect{I}_S ,
\end{align}
where $\vect{D}_l$ is given as
\begin{align}
\vect{D}_l = \begin{bmatrix} d_{l,0,0} & \ldots & d_{l,0,S-1} \\ \vdots & \ddots & \vdots \\ d_{l,S-1,0} & \ldots & d_{l,S-1,S-1} \end{bmatrix}.
\end{align}
Expressing all the subcarrier channels in \eqref{eq:frequency} in vector form, we write
\begin{align}
\underbrace{\begin{bmatrix}\bar{y}[0]  \\ \vdots \\ \bar{y}[S-1] \end{bmatrix}}_{=\bar{\vect{y}}} =\underbrace{\diag\left(\bar{h}[0],\ldots,\bar{h}[S-1]\right)}_{=\bar{\vect{H}}} \underbrace{\begin{bmatrix}\bar{\chi}[0] \\  \vdots \\ \bar{\chi}[S-1]\end{bmatrix}}_{=\bar{\boldsymbol{\chi}}} + \underbrace{\begin{bmatrix}\bar{n}[0] \\ \vdots \\ \bar{n}[S-1]\end{bmatrix}}_{=\bar{\vect{n}}}.  \end{align}
Denoting the $S\times S$ scaled discrete Fourier transform (DFT) matrix (scaled to have unitary matrix) by $\vect{F}_S\in \mathbb{C}^{S\times S}$, the frequency-domain noise vector $\bar{\vect{n}}$ is distributed as $\bar{\vect{n}}\sim\CN(\vect{0},\vect{F}_S\vect{D}\vect{F}_S^{\Htran})$. Denoting $\bar{\vect{D}}=\vect{F}_S\vect{D}\vect{F}_S^{\Htran}$, we first apply whitening to the received signal $\bar{\vect{y}}$ and obtain
\begin{align}
\bar{\vect{y}}'=\bar{\vect{D}}^{-1/2}\bar{\vect{y}}=\underbrace{\bar{\vect{D}}^{-1/2}\bar{\vect{H}}}_{\triangleq \bar{\vect{H}}'}\bar{\boldsymbol{\chi}}+\underbrace{\bar{\vect{D}}^{-1/2}\bar{\vect{n}}}_{\triangleq \bar{\vect{n}}'}
\end{align}
where $\bar{\vect{n}}'\sim \CN(\vect{0},\vect{I}_S)$. Denoting the singular value decomposition of the effective channel as $\bar{\vect{H}}'=\vect{U}\vect{\Sigma}\vect{V}^{\Htran}$, the capacity of the repeater-assisted OFDM channel is obtained as \cite{bjornson2024introduction}
\begin{align}
C = \frac{B}{T+S} \sum_{\nu=0}^{S-1}\log_2\left(1+\sigma_\nu^2q_{\nu}\right), \quad \text{bit/s},
\end{align}
where $\sigma_{0}\geq \sigma_{1} \geq \ldots \geq \sigma_{S-1}$ are the ordered singular values along the diagonal entries of $\vect{\Sigma}$. The capacity-achieving power allocation coefficients, $q_\nu$ are found by water-filling algorithm under the total power constraint $\sum_{\nu=0}^{S-1}q_{\nu}=qS$ (here, $q$ is the average power per time-domain symbol). The capacity-achieving input distribution is given as $\bar{\boldsymbol{\chi}}\sim \CN(\vect{0}, \vect{V}\diag(q_0,q_1,\ldots,q_{S-1})\vect{V}^{\Htran})$.

\section{Numerical Results}
In this section, we compare the capacities achieved over a repeater-assisted SISO-OFDM channel under different repeater activation schemes. We consider a $1$\,km$\times$1\,km network area with a BS located at the center, and $L=16$ repeaters deployed on a uniform $4\times 4$ grid. The results are averaged over 25 random UE placements, where in each realization the UE is dropped uniformly at random within the network area. The heights of the BS, repeaters, and UE are set to $25$\,m, $15$\,m, and $1.5$\,m, respectively. We assume line-of-sight (LOS) channels with multiple reflected paths to and from the repeaters, while the direct channel between the BS and UE is non-line-of-sight (NLOS). The carrier frequency is $3$\,GHz, and the channel modeling follows the 3GPP guidelines in \cite{3GPP25996}, adopting the same channel generation mechanism as in \cite[Fig.~9.19]{bjornson2024introduction}. The capacity is computed as the average over 10 random realizations of the multipath components for each UE placement.  

We investigate four configurations. The first is \emph{No repeaters}, where only the direct BS--UE channel contributes to the communication. The second is \emph{All repeaters}, in which all repeaters are simultaneously activated with the same amplification factor $\alpha$. The third configuration, \emph{One repeater}, activates only the repeater closest to the UE, while all others remain inactive. Finally, the \emph{Closeby+Rand} configuration activates the closest repeater together with three additional repeaters chosen at random. The latter two cases are included to assess whether activating all repeaters is essential or if a reduced subset may already yield competitive performance.

\begin{figure}[t!]
        \centering
	\begin{overpic}[width=0.98\columnwidth,trim=0.2cm 0cm 0.5cm 0.5cm,clip,tics=10]{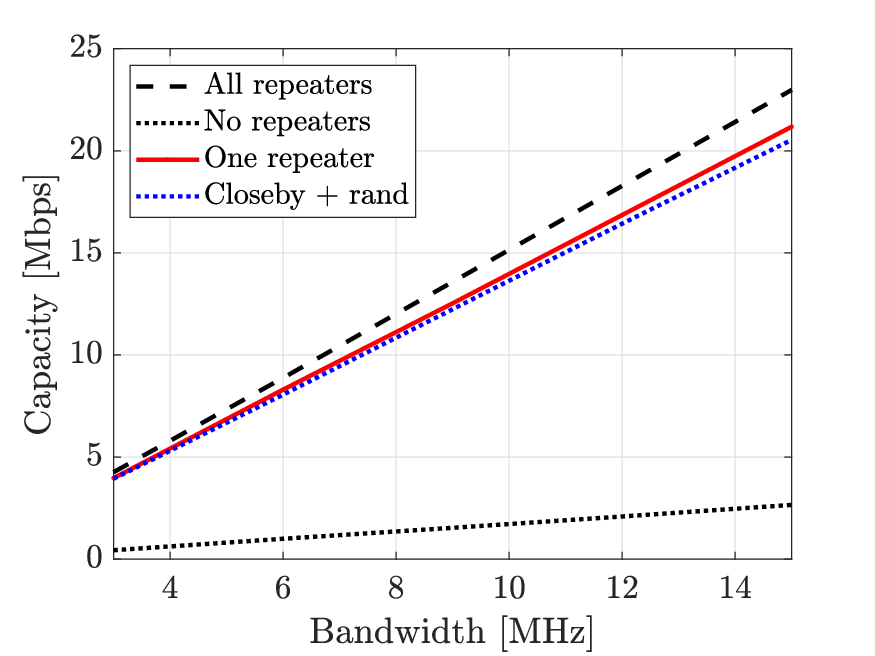}
\end{overpic} 
\vspace{-4mm}
        \caption{The capacity in terms of total bandwidth. }
        \label{fig1}
        \vspace{-2mm}
\end{figure}

In Fig.~\ref{fig1}, we set $\alpha=30$\,dB and present the capacity as a function of the bandwidth $B$, assuming that the transmit power scales proportionally with $B$ to maintain a signal power spectral density of $20$\,mW/MHz. The subcarrier spacing is $150$\,kHz, which increases the number of subcarriers and channel taps with the bandwidth, while the multipath environment remains unchanged. As observed in the figure, activating even a single repeater (the closest repeater) yields a significant capacity improvement compared to the case without repeater assistance. This gain becomes even more pronounced as the bandwidth increases. Interestingly, the performance gap between the \emph{All repeaters} and \emph{One repeater} configurations remains small, indicating that activating only the closest repeater can provide substantial energy savings compared to activating all repeaters. Of course, this conclusion holds for a single-UE setup, while the extension to multi-UE scenarios remains an open research question. Moreover, when the closest repeater is activated together with three randomly selected repeaters, we observe a performance degradation. This highlights the critical importance of carefully selecting which repeaters to activate, making it an appealing optimization problem for future investigation.

\begin{figure}[t!]
        \centering
	\begin{overpic}[width=0.98\columnwidth,trim=0.1cm 0cm 0.5cm 0.5cm,clip,tics=10]{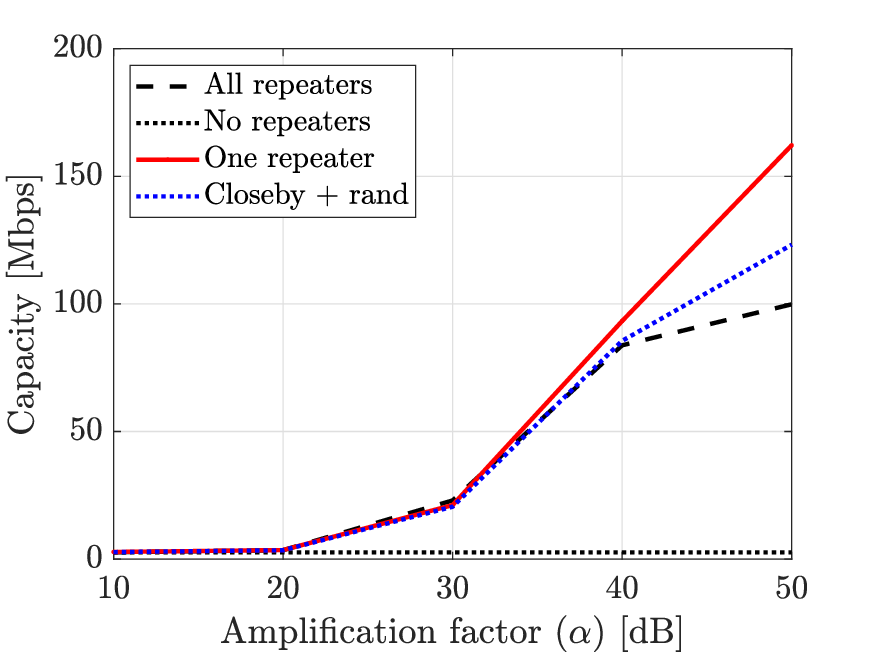}
\end{overpic} 
\vspace{-4mm}
        \caption{The capacity in terms of amplification factor $\alpha$. }
        \label{fig2}
        \vspace{-2mm}
\end{figure}

In Fig.~\ref{fig2}, we fix the number of subcarriers to $1000$ ($B=15$\,MHz) and vary the amplification factor of the repeaters. When the amplification factor is too small, repeater assistance provides no benefit. However, once it exceeds $20$\,dB, a significant capacity gain is observed. Beyond a certain point, activating only the closest repeater yields the highest capacity, while the \emph{All repeaters} configuration saturates, since the distant repeaters contribute primarily to noise amplification rather than improving the signal-to-noise ratio. This effect is due to the inherent noise amplification mechanism of the repeaters. An important aspect not addressed in this study is the total transmit power budget of the repeaters. The joint optimization of the amplification factors across different repeaters, together with power constraints, is left for future work.

\section{Conclusions}
In this paper, we analyzed the capacity of wideband SISO-OFDM systems assisted by a swarm of NCRs. Starting from a continuous-time passband model, we derived a rigorous capacity expression that accounts for frequency-selective propagation and noise amplification effects. Our numerical results demonstrated that NCRs can substantially enhance system capacity even with simple activation strategies. In particular, activating only the closest repeater yields almost the same performance as activating all repeaters, thereby enabling significant energy savings. Conversely, activating additional distant repeaters may even degrade performance due to noise amplification, highlighting the importance of repeater selection and amplification control. These findings establish NCR swarms as a cost-effective and scalable solution for coverage extension and capacity improvement in future wideband wireless systems. An important direction for future work is the joint optimization of repeater activation and amplification factors under realistic power constraints and multi-user scenarios.

\bibliographystyle{IEEEbib}
\bibliography{strings,refs}

\end{document}